\RequirePackage{etoolbox}
\csdef{input@path}{%
 {sty/}
 {img/}
}%
\csgdef{bibdir}{bib/}

\documentclass[]{imsart}

\pubyear{0000}
\volume{00}
\issue{0}
\doi{0000}
\firstpage{1}
\lastpage{1}

\RequirePackage[OT1]{fontenc}
\RequirePackage[numbers]{natbib}
\RequirePackage[colorlinks,citecolor=blue,urlcolor=blue,filecolor=blue,backref=page]{hyperref}
\RequirePackage{amsthm,amssymb,mathrsfs,graphicx,graphics,algorithm,algorithmic}
\RequirePackage{amsmath}

\RequirePackage{geometry}
\RequirePackage{subfigure}
\RequirePackage{epsfig}
\RequirePackage{lipsum} 
\allowdisplaybreaks

\startlocaldefs
\numberwithin{equation}{section}
\theoremstyle{plain}

\endlocaldefs

\begin{document}

\begin{frontmatter}
\title{Some variations on Ensembled Random Survival Forest with application to Cancer Research}
\runtitle{}

\begin{aug}
\author{\fnms{Arabin K.} \snm{Dey}\thanksref{}\ead[label=e1]{arabin@iitg.ac.in}}\\
\affiliation[]{Department of Mathematics, IIT Guwahati} \\ \printead{e1}\\
\author{\fnms{Suhas} \snm{N.}\thanksref{}\ead[label=e2]{suhas.n@iitg.ac.in}},\\
\affiliation[]{Department of Mathematics, IIT Guwahati} \\ \printead{e2}\\
\author{\fnms{Talasila} \snm{Sai Teja}\thanksref{}\ead[label=e3]{talasila@iitg.ac.in}}\\
\affiliation[]{Department of Mathematics, IIT Guwahati} \\ \printead{e3}\\
\author{\fnms{Anshul} \snm{Juneja}\thanksref{}\ead[label=e4]{anshuljuneja13.aj@iitg.ac.in}}\\
\affiliation[]{Department of Mathematics, IIT Guwahati} \\ \printead{e4}\\

\runauthor{Dey Suhas Talasila and Juneja}


\end{aug}

\begin{abstract}
   In this paper we describe a novel implementation of adaboost for prediction of survival function.  We take different variations of the algorithm and compare the algorithms based on system run time and root mean square error.  Our construction includes right censoring data and competing risk data too.  We take different data set to illustrate the performance of the algorithms.    
\end{abstract}



\end{frontmatter}

\section{Introduction}

 Random survival forest is a very popular tree-based method in predicting survival function given a set of covariates.  It is a widely used algorithm in biomedical research.  In this paper we introduce different variations of ensembled Random Survival Forest and compare them with few existing variations as an application of predicting survival of a cancer patient.  One popular tree based ensemble is Adaboost.  It takes many weak learners additively to form a strong learner.  We apply them in survival prediction.  Our algorithms work in right censoring set up too.  Modeling survival function depending on cause of the failure is an interesting dimension.  This is known as competing risk models.  We also extend our ensembled methods for predicting survival function in case of competing risk set up. 

   In early eighties idea of randomized decision appeared in experimental studies.   During early nineties statistical notion of variance and bias in tree models was studied by some researchers (\cite{Breiman:1996a}, \cite{Friedman:1997}). Breiman \cite{Breiman:1996b} came up in 1996 with the "Bagging" idea in order to reduce variance of a learning algorithm without increasing its bias too much.  Several generic randomization methods have been proposed, like bagging, are applicable to any machine learning algorithm.  In this connection, Geurts et. al. \cite{GeurtsErnstWehenkel:2006} introduced extremely randomized trees as a variation of RF which permits to build a totally randomized random trees.  Recently Adaboost technique has become an attractive ensemble method in machine learning [\cite{ThongkamXuZhang:2008}, \cite{MaDing:2003}].  In this paper we describe \textbf{a novel structure of Adaboost algorithm} which can also be used for certain other type of regression problem.  The decision problem are also used in prediction of survival function. Early experimental work by Breiman \cite{Breiman:2010} on survival forests is also relevant.  In this approach a survival tree is grown using a hybrid splitting method in which nodes are split both on time and covariates.  Recently Ishawaran et. al. \cite{IshwaranKogalurBlackstoneLauer:2008} proposed random survival forest (RSF) for analysis of right-censored survival data.  But there is no work/methodology for computing conditional survival function by extending RSF or its variations in ensembled set up like Adaboost.    

 We organize the paper in the following way.  In section 2, we provide the conventional structure of random survival forest.  Extra survival tree and Adaboost with extra survival tree are provided in section 3.  Random Survival Tree and Extra Survival Trees under Competing Risk data is discussed in section 4.  Data analysis is kept in section 5.  We finally conclude in section 6.

\section{Usual Structure of Random Survival forest}

 Like Random forest, Random survival forest builds many binary trees, but aggregation scheme is now based on a cumulative hazard function (CHF) described in more details below.

Steps of the random forest can be given as follows :

\begin{enumerate}

\item Draw bootstrap samples from the original data ntree times.  For each bootstrap sample, this leaves approximately one-third of the samples out-of-bags (OOB).

\item A survival tree is grown for each bootstrap sample.

\begin{enumerate}

\item At each node of the tree, select $\sqrt{n}$ predictors at random for splitting.

\item Using one of the splitting criteria described below, a node is split using the single predictor from step-II(a) that maximizes the survival differences between daughter nodes. 

\item Repeat steps (II(a) and b) until each terminal node contains no more than 0.632 times the number of events.

\end{enumerate}

\item Calculate a CHF for each survival tree built.  Aggregate the ntree trees to obtain the ensemble cumulative hazard estimate. 

\end{enumerate}

\subsection{Split Criteria :}  Although there are four major split criteria is available in the literature, we use log-rank based split criteria for our purpose.  LR test for splitting is defined as follows :
\begin{eqnarray*}
L(X, c) = \frac{ \sum_{i = 1}^{N} d_{t_{i}, child_{i}} - R_{t_{i}, child_{1}}\frac{d_{t_{i}}}{R_{t_{i}}} }{\sqrt{\sum_{i = 1}^{E} \frac{d_{t_{i}}(R_{t_{i}} - d_{t_{i}})}{R_{t_{i}} - 1}\frac{R_{t_{i}, child_{1}}}{R_{t_{i}}}(1 - \frac{R_{t_{i}, child_{1}}}{R_{t_{i}}})}}
\end{eqnarray*}

where, N is the number of distinct event times $T_{(1)} \leq T_{(2)} \leq \cdots \leq T_{(N)}$ in the parent node, $d_{t_{i}, child_{j}}$ is the number of events at time $t_{i}$ in the child nodes, $j = 1, 2$, $R_{t_{i}, child_{j}}$ is the number of individuals at risk at time $t_{i}$ in the child nodes $j = 1, 2, \cdots$ i.e. then number of individuals who are alive or dead at time $t_{i}$, and $R_{t_{i}} = R_{t_{i}, child_{1}} + R_{t_{i}, child_{2}}$ and $d_{t_{i}} = d_{t_{i}, child_{1}} + d_{t_{i}, child_{2}}$.  The absolute value of $LR(X, c)$ measures the node separation.  The best split is chosen in such a way that it maximizes the absolute value of 
\begin{eqnarray*}
LRS(X, c) =  \frac{\sum_{x_{i} \leq c}^{} a_{i} - n\mu_{a}}{\sqrt{n_{1}(1 - \frac{n_{1}}{n}) s^{2}_{a}}}
\end{eqnarray*}

where, $\mu_{a}$ and $s^{2}_{a}$ are sample mean and sample variance of $a_{i}$, respectively.  LRS(X, c) measures node separation.

\subsection{Ensemble CHF} Once survival reaches step-III in the algorithm, i.e. until each terminal node contain no more than 0.632 times the number of events, the trees are aggregated to form an ensemble CHF, which is calculated by grouping hazard estimates using terminal nodes.  Let $L$ be a terminal node, $t_{i, L}$ be distinct survival times, $d_{t_{i, L}}$ be the number of events and $R_{t_{i, L}}$ be the individual at risk at time ($t_{i, L}$).

 The CHF estimate for a terminal node L is the Nelsen-Aalen estimator.  $$ \hat{\Lambda}_{L}(t) = \sum_{t_{i, L} \leq t}^{} \frac{d_{t_{i}, L}}{R_{t_{i}, L}}. $$  All individual within L will have same CHF.  For $q$ terminal nodes in a tree, there are $q$ different CHF values.  To determine $\hat{\Lambda}_{L}(t)$ for an individual $i$ with predictor $x_{new}$, drop $x_{new}$ down the tree and the $x_{new}$ will fall into a unique terminal node $L \in Q$.  CHF at L would be the CHF for individual i in the test sample.  The bootstrap ensemble for individual i is $$ \Lambda^{*}(t | x_{new}) = \frac{1}{ntree}\sum_{b = 1}^{ntree} \Lambda^{*}_{b}(t | x_{new}), $$ where $\Lambda^{*}_{b}(t | x_{new})$ is CHF for a particular tree.  For prediction, ensemble survival is defined as $$ S(t | x_{new}) = e^{-\Lambda^{*}(t | x_{new})} $$

\subsection{Extra Survival Trees :} A slightly different version of random survival forest is available on extra tree approach [\cite{IshwaranKogalurBlackstoneLauer:2008}].  We call it as Extra Survival Forest (ESF).  Therefore algorithmic steps for ESF are as follows :

\begin{enumerate}

\item A survival tree is grown for each sample.

\begin{enumerate}

\item At each node of the tree, select $\sqrt{n}$ predictors at random for splitting.

\item Select set of random points to exercise splitting criteria from each predictors. 

\item Using Log-rank splitting criteria described in previous section, a node is split using the single predictor that maximizes the survival differences between daughter nodes. 

\item Repeat steps (II(a), (b) and (c)) until each terminal node contains no more than 0.632 times the number of events.

\end{enumerate}

\item Calculate a CHF for each survival tree built.  Aggregate the ntree trees to obtain the ensemble cumulative hazard estimate. 

\end{enumerate}

\section{Our Proposed Variations}

  We propose to use Adaboost with different weak learner as survival forests.  Adaboost is most popular ensemble method.  Generally it is used for prediction in classification tasks.  In this paper we use adaboost as a regression problem. Recently Kannao and Guha \cite{KannaoGuha:2015} used a ensembled method based on weighted mean square error as loss function.  Our formulation is based on a slightly different exponential loss function than what usual adaboost algorithm does.  
 
\subsection{Mathematical Foundation of the Proposed Algorithm}

  Similar to the case of minimizing exponential error in Adaboost algorithm, we now show that the proposed Adaboost with Survival Forest algorithm reduces the exponential error.

 We define the exponential error in our case as: $ E = \sum_{n=1}^{N} e^{-I_m(x_n)}. $
where $I_m$ is a classifier defined as follows $ I_m = sign(\frac{1}{2}\sum_{l=1}^{m}\alpha_l I(y_l(x_n), t_n)). $
where $y_i$'s are independent forests used as regressors and 
Therefore,
\begin{eqnarray*}
E & = & \sum_{n=1}^{N} e^{-\frac{1}{2}\sum_{l=1}^{m}\alpha_l I(y_l(x_n) \neq t_n)}\\
 & = & \sum_{n=1}^{N} e^{-\frac{1}{2}\alpha_m I(y_m(x_n) \neq t_n)} \times e^{-\frac{1}{2}\sum_{l=1}^{m-1}\alpha_l I(y_l(x_n)\neq t_n)}\\ & = & \sum_{n=1}^{N} w_{n}^{(m)}e^{-\frac{1}{2}\alpha_m I(y_m(x_n) \neq t_n)}
\end{eqnarray*}

where we take $w_{n}^{(m)}=\sum_{l=1}^{m-1}\alpha_l I(y_l(x_n)\neq t_n))$ to be constants as we assume the regressors $y_1, y_2, \cdots, y_
{m-1}$ to be fixed.

\begin{eqnarray}
E = e^{-\frac{\alpha_m}{2}}\sum_{T_m}w_{n}^{(m)} + e^{\frac{\alpha_m}{2}}\sum_{T_{m}^{c}}w_{n}^{(m)}
\end{eqnarray}

where $T_m$ is set of data points that are classified by $y_m(x)$ and we denote the misclassified points by $T_{m}^{c}$ 

\begin{eqnarray}
E = (e^{\frac{\alpha_m}{2}}-e^{-\frac{\alpha_m}{2}})\sum_{n=1}^{N}w_{n}^{(m)}I(y_m(x_n)\neq t_n) + e^{-\frac{\alpha_m}{2}}\sum_{n=1}^{N}w_{n}^{(m)}
\end{eqnarray}
 
 Note that $I(y_m(x_n)\neq t_n)$ in our case represents points which are misclassified when the underlying regressor $y_m$ produces error.
Minimizing in terms of $y_m(x)$ the second term is a constant and so minimization of the first term occurs as we perform Adaboost.

 We then form the strong regressor using the weights $\alpha$ generated in the above method. That is, $$Y_m(x_n)=\sum_{l=1}^{m}\alpha_l y_l(x_n).$$



\subsection{Adaboost on Survival Trees :}  The algorithmic steps for this generic adaboost algorithm for survival forest can be given by
 
\begin{enumerate}

\item Input : S : training set, $S = x_{i}$ ($i = 1, 2, \cdots, n$), labels $y_{i} \in Y$, $k :$ Iterations number.

\item for k = 1(1) K
     \begin{enumerate}
       \item Draw random sample of size n from S with weights $w_{i}$.
       \item Fit a regressor $y_{m}(x)$ to the training data by minimizing the weighted error function : $$ J_{m} = \sum_{n = 1}^{N} w^{m}_{n} I(y_{m}(x_{n}) \neq t_{n}) $$ where, $I(y_{m}(x_{n}) \neq t_{n})$ is the indicator function and equals to -1 when $y_{m}(x_{n}) \neq t_{n}$ and 1 otherwise.
       \item Evaluate the quantities $$ \epsilon_{m} = \frac{\sum_{n = 1}^{N}w^{m}_{n}I(y_{m}(x_{n}) \neq t_{n})}{\sum_{n = 1}^{N} w^(m)_{n}} $$  and then use these to evaluate $\alpha_{m} = \ln(\frac{(1 - \epsilon_{m})}{\epsilon_{m}})$

       \item Update the data weighting co-efficients $$ w^{m + 1}_{n} = w^{m}_{n} \exp(\alpha_{m}I(y_{m}(x_{n}) \neq t_{n})) $$

     \end{enumerate}

\item Make predictions using the final model, which is given by $$ Y_{M}(x) = \sum_{m = 1}^{M} \alpha_{m} y_{m}(x). $$

\end{enumerate}

 Each of the terminal node provides set of survival times based on which cumulative hazard is calculated.  Instead of calculating cumulative hazard at each tree and combined them, what we do is as follows 

\noindent{\textbf{Variation 1: Mean of Mode} :} For each classifier/regressor, we extract mode of unique survival times at each tree.  Finally we predict the mean of those survival time instead of calculating CHF at each terminal node.  The method can be explored in censored data set too.    

\noindent{\textbf{Variation 2: Mapped Mean of Mode} :}  In our context we adapt the last two methods as they provide better approximation of the survival. However after calculating mean in variation 1, we see what are those unique time points in training example that contain mean.  We consider the final predicted survival time as that unique survival time which is nearest to the mean. 

\subsection{Proposed Variation in Above context :}  

 In the paper the identity function taken as a classifier formed based on a certain type of regressors.  

\noindent{\textbf{Variation in regressor 1:} }  In the case we take all independent regressor in adaboost as random survival forest. We call this variation as ADARSF.

\noindent{\textbf{Variation in regressor 2:}} Here we take all independent regressor in adaboost as Extra survival forest. We call this variation as ADAESF.

\noindent{\textbf{Variation in regressor 3:}} Here we take all regressor in adaboost as mixture of Random Survival Forest and Extra Survival Forest.  We call this variation as ADAMIX.    
   
\section{Different Survival Trees under Competing Risk Data}

 In survival and medical studies it is quite common that more than one cause of failure may be directed to a system at the same time.  It is often interesting that an investigator needs to estimate a specific risk in presence of other risk factors.  In statistical literature such risk is known as competing risk model.  Iswaran et al. \cite{IshwaranGerdsKogalurMooreGangeLau:2014} explored random survival forest in case of competing risk model. In this paper we may like to explore our proposed variations in a competing risk data set where data set can be Type-I censored or not censored.  Competing Risks are sometimes treated little differently which depends on primary end points where it should be relapse free survival or relapse itself.  To make detailed analysis we consider different data sets and carry out analysis. 

\subsection{Cause-specific Estimations :}  In this section we discuss cause-specific survival function and cause specific cumulative hazard function.  Let us assume the following  :

\begin{enumerate}

\item $t_{1} < t_{2} < \cdots < t_{m}$ denote $m < n$ distinct and ordered time to events from $(T_{i})_{1 \leq i \leq m}$

\item $d_{j}(t_{k}) = \sum_{i = 1}^{n} I(T_{i} = t_{k}, \delta_{i} = j)$ be the count of the events of type $j$ that have happened at time point $t_{k}$.

\item $N_{j}(t) = \sum_{i = 1}^{n} I(T_{i} \leq t_{k}, \delta_{i} = j)$ be the number of type $j$ events that happen in time interval $[0, t_{k}]$.

\item $d(t_{k}) = \sum_{j}^{} \delta_{j}(t_{k})$ be the total number of failures occuring in the interval $[0, t]$. 

\item $N(t) = \sum_{j}^{} N_{j}(t)$ be the total number of failures occuring in the interval $[0, t]$.

\item $Y(t) = \sum_{i = 1}^{n} I(T_{i} \leq t)$ be the total count of individuals which are at risk (both event-free and uncensored) just prior to time point t.  

\end{enumerate}

 The Nelson-Aalen estimator for the cause-specific cumulative hazard function defined by $\Lambda_{j}(t) = E_{X}[\int_{0}^{t} \alpha_{j}(s|X) ds]$ is given by $$ \hat{\Lambda}_{j}(t) = \int_{0}^{t} \frac{dN_{j}(s)}{Y(s)} ds = \sum_{k = 1}^{m(t)} \frac{d_{j}(t_{k})}{Y(t_{k})} $$ where, $m(t) = \max\{k : t_{k} \leq t \}$.

 The Kaplan-Meier estimator for the event-free survival function is given by $$ \hat{S}(t) = \prod_{s \leq t}^{} (1 - \frac{N(ds)}{Y(s)}) = \prod_{k = 1}^{m(t)} (1 - \frac{d(t_{k})}{Y(t_{k})}). $$  We use Aalen-Johansen estimator (\cite{Borgan:2005}) to estimate cause-specific survival function $F_{j}(t)$ for $j$-th cause : 
$$ \hat{F}_{j}(t) = \int_{0}^{t} \hat{S}(u-)d\hat{H}_{j}(u) = \int_{0}^{t} \hat{S}(u-)Y^{-1}(u)N_{j}(du) = \sum_{k = 0}^{m(t)} \hat{S}(t_{k-1})Y^{-1}(t_{k})d_{j}(t_{k}). $$

 Algorithm for competing risk in case of straight random survival forest carry the following steps :  

\begin{enumerate}

\item Draw B bootstrap samples from the learning data by replacement.  

\item Grow a competing risk tree for each of the bootstrap samples.  Randomly select mtry predictor variables at every node of the tree.  Choose the predictor variable that maximizes the competing risk splitting rule.  

\item Continue tree growing as long as the number of non-censored observations in each node is larger than a pre-specified minimum terminal node size, termed node-size.

\item Calculate a cause-specific predicted survival time i.e. mode of unique survival times at each tree.  Average/mapped version of the average to obtain the ensemble estimates.

\end{enumerate}

 We propose a similar variation in the similar line using extra survival tree.  The algorithm is as follows :

\begin{enumerate}

\item Take all training sample corresponding to a particular cause.

\item To grow a competing risk survival tree, we randomly select p candidate variables at each node of the tree.

\item Take a random split at each selected covariate to divide the data set into two groups (which will act as two daughter nodes).  The node is split using the candidate variable that maximizes the cause-specific survival difference between daughter nodes.

\item Grow the tree to full size under the constraint that a terminal node should have no less than $d_{0} > 0$ unique deaths.

\item Calculate a cause-specific survival time i.e. mode of unique survival times at each tree.  Average/mapped version of the average to obtain the ensemble CHF or survival function.   

\end{enumerate}

\section{Data Analysis}  We implement the algorithms in three different data sets.  The codes are written using R - 3.3.2. All codes can be available from authors by request.  The programs are run on Intel(R) Xeon(R) server with processor core 4 CPU E5620 @ 2.4GHz at Department of Mathematics IIT Guwahati.      

\subsection{Data Set 1 (Prediction of Survival Without Competing Risk):}  This breast cancer dataset contains gene expression and clinical data published in Desmedt et al. \cite{Desmedt:2007} The data contains 198 samples to independently validate a 76 gene prognostic breast cancer signature as part of the TransBig project.  In the data, 22283 gene features and 21 clinical covariates are provided for each sample.  The dataset can be obtained through the R package "breastCancerTRANSBIG" of "Bioconductor".     

\begin{figure}
 \begin{center}
  \includegraphics[width = 1\textwidth]{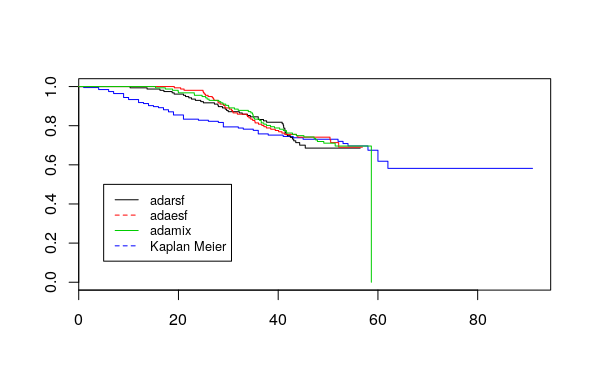} 
\caption{TRANSBIG Data Set (Mapped mean of mode): Survival functions provided by different methods \label{SFDM1}}
\end{center}
\end{figure}

\begin{figure}
 \begin{center}
  \includegraphics[width = 1\textwidth]{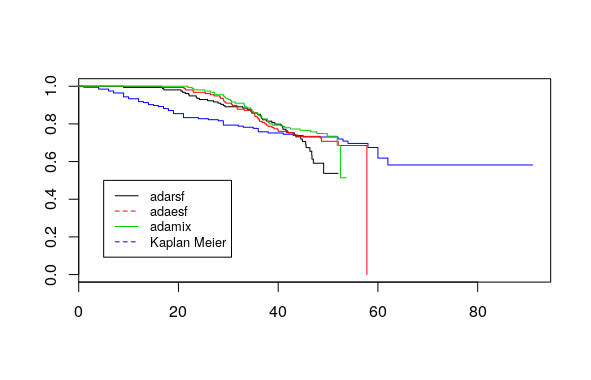} 
\caption{TRANSBIG Data Set (Mean of mode): Survival functions provided by different methods \label{SFDM2}}
\end{center}
\end{figure}

 We predict overall survival time based on other covariates.  We use Adaboost with random survival forest and extra survival trees as weak learners.  We also compare them with Kaplan Meier estimates. The survival functions based on above methods are shown in Figure-\ref{SFDM1} and Figure-\ref{SFDM2}.  Figure-\ref{SFDM1} uses mapped mean of mode method in ensemble while Figure-\ref{SFDM2} uses straight mean of mode.  Results for Adaboost with RSF and ESF with Mean of mode of terminal node as event time is performed based on 10 trees. Mean of modes and mapped mean of modes can be found in Table-\ref{Tab-transbig-1} and Table-\ref{Tab-transbig-2} respectively.  

\textbf{Overall Comment :}  RMSEs can be improved by growing more trees.  System run time is on higher side as dimension of the data is very high.  Performance of ADA-ESF seems moderately well or best across all the methods.


\begin{table}
        \centering
        \begin{tabular}{ |c|c|c|c|}
            \hline
            \multicolumn{3}{|c|}{RMSE} & \\
            \hline
             & train &test & Running Time(sec)\\ 
            \hline
             ADA-RSF & 16.02 & 16.40 & 51.85\\ 
             ADA-ESF & 12.74 & 19.64 & 50.46\\ 
             ADA-MIX & 12.89 & 21.43 & 50.77\\ 
             \hline
        \end{tabular}
        \caption{Data Set 1 : Mean of modes - RMSE 10 iterations each with 10 trees \label{Tab-transbig-1}}        
\end{table}

\begin{table}
        \centering
        \begin{tabular}{ |c|c|c|c|}
            \hline
            \multicolumn{3}{|c|}{RMSE} & \\
            \hline
             & train &test & Running Time(sec)\\ 
            \hline
             ADA-RSF & 13.50 & 20.74 & 48.04\\ 
             ADA-ESF & 13.14 & 15.50 & 54.38\\ 
             ADA-MIX & 13.26 & 18.73 & 51.05\\ 
             \hline
        \end{tabular}
        \caption{Data Set 1 : Mapped mean of modes - RMSE 10 iterations each with 10 trees \label{Tab-transbig-2}}        
\end{table}

\subsection{Data Set 2 in Competing Risk set up : Application on Follicular Cell Lymphoma Data}  We consider follicular cell lymphoma data from Pintilie \cite{Pintilie:2006} where additional details about data set can be found. The data set can be downloaded from \url{https://www.jstatsoft.org/article/view/v38i02}, and consists of 541 patients with early disease stage follicular cell lymphoma (I or II) and treated with radiation only (chemo = 0) or a combined treatment with radiation and chemotherapy (chemo = 1). Parameters recorded were path1, ldh, clinstg, blktxcat, relsite, chrt, survtime, stat, dftime, dfcens, resp and stnum. The two competing risks are \textbf{death without relapse} and \textbf{no treatment response}. The patient's ages (age: mean = 57 and sd = 14) and haemoglobin levels (hgb: mean = 138 and sd = 15) were also recorded. The median follow-up time was 5.5 years. There are more parameters which are not of our concern.

  First we preprocess the data to find the cause of failure indicator. There are 272 (no treatment response or relapse) events due to the disease, 76 competing risk events (death without relapse) and 193 censored individuals. The event times are denoted as dftime. Thus our data set is prepared.  We plot cause specific survival function based on Kaplan Meier and Adaboost with its different variations.  Plots are available in Figure-\ref{SUR_FOLLIC}.



\begin{figure}
\begin{center}
 \subfigure[$\xi_{1}$]{\includegraphics[width = 0.5\textwidth]{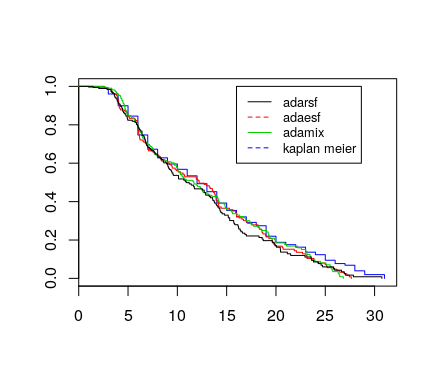}}  
 \subfigure[$\xi_{2}$]{\includegraphics[width = 0.5\textwidth]{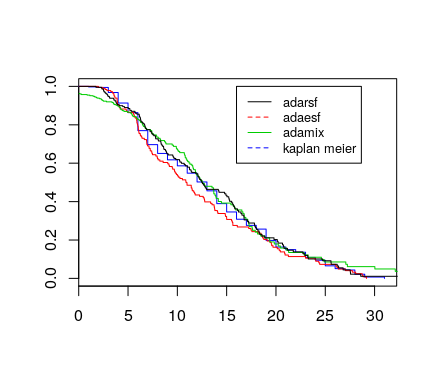}} \\
\caption{Mean of Mode  for Follic Data : Survival Function based on Kaplan Meier and different variations of Survival Trees \label{Sur_Follic}}
\end{center}
\end{figure}

\begin{figure}
\begin{center}
 \subfigure[$\xi_{1}$]{\includegraphics[width = 0.5\textwidth]{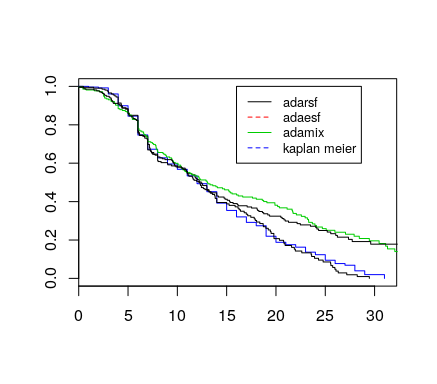}}  
 \subfigure[$\xi_{2}$]{\includegraphics[width = 0.5\textwidth]{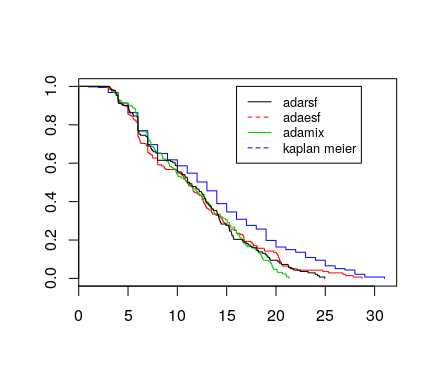}} \\
\caption{Mapped Mean of mode for Follic Data : Survival Function based on Kaplan Meier and different variations of Survival Trees \label{Sur_Follic1}}
\end{center}
\end{figure}

 Results for two competing causes (Follic death and relapse) can be found in Table-\ref{FOLDEATH} and Table-\ref{FOLRELAPSE} respectively,  while performing adaboost with RSF and ESF with mean of mode in terminal node as event time.  Both runtime and test error is large for Follic relapse case. Results for mapped mean of mode are available at Table-\ref{Follic-death-mapped-mean-modes} and Table-\ref{Follic-relapse-mapped-mean-modes} respectively.  Figure-\ref{Sur_Follic} and Figure-\ref{Sur_Follic1} shows the plots for cause specific survival function for Mean of mode and Mapped mean of mode when different regressors are applied.  

\textbf{Overall Comment :}  It is difficult to predict one particular best method across all the cases.  However Figures show that all the methods are capable of predicting the survival curve.  ADA-ESF seems carrying a moderate or best RMSEs and system run times across all cases.

\begin{table}
        \centering
        \begin{tabular}{ |c|c|c|c| }
            \hline
            \multicolumn{3}{|c|}{RMSE} & Running Time(sec) \\
            \hline
             ADA-ESF & 1.829 & 1.536 & 0.75\\ 
             ADA-MIX & 1.329 & 1.437 & 0.79\\ 
             ADA-RSF & 1.801 & 2.043 & 0.64\\ 
             \hline
        \end{tabular}
        \caption{Follic death case : Mean of modes - Running Time 10 iterations each with 10 trees \label{FOLDEATH}}        
\end{table}

\begin{table}
        \centering
        \begin{tabular}{ |c|c|c|c| }
            \hline
            \multicolumn{3}{|c|}{RMSE} & \\
            \hline
             & train & test & Running Time(sec)\\ 
            \hline
             ADA-ESF & 2.09 & 2.27 & 1.06\\ 
             ADA-MIX & 7.28 & 8.27 & 0.90\\ 
             ADA-RSF & 1.99 & 2.31 & 0.70\\ 
             \hline
        \end{tabular}
        \caption{Follic relapse case : Mean of modes - RMSE 10 iterations each with 10 trees \label{FOLRELAPSE}}
\end{table}

\begin{table}
        \centering
        \begin{tabular}{ |c|c|c|c| }
            \hline
            \multicolumn{3}{|c|}{RMSE} & \\
            \hline
             & train & test & Running Time(sec)\\ 
            \hline
             ADA-ESF & 1.66 & 1.71 & 0.88\\ 
             ADA-MIX & 9.30 & 12.55 & 0.80\\ 
             ADA-RSF & 7.74 & 9.21 & 4.19\\ 
             \hline
        \end{tabular}
        \caption{Follic death case  : Mapped mean of modes - RMSE in 10 iterations each with 10 trees \label{Follic-death-mapped-mean-modes}}        
\end{table}

\begin{table}
        \centering
        \begin{tabular}{ |c|c|c|c| }
            \hline
            \multicolumn{3}{|c|}{RMSE}& \\
            \hline
             & train &test& Running Time(sec)\\ 
            \hline
             ADA-ESF & 3.22 & 3.79 & 1.20\\ 
             ADA-MIX & 2.36 & 3.17 & 1.08\\ 
             ADA-RSF & 2.00 & 1.88 & 4.47\\ 
             \hline
        \end{tabular}
        \caption{Follic relapse case : Mapped mean of modes - RMSE 10 iterations each with 10 trees \label{Follic-relapse-mapped-mean-modes}}        
\end{table}

\subsection{Data Set 3 :  Mayo Clinic trial in Primary Biliary Cirrhosis (PBC) :}  
\begin{figure}
\begin{center}
 \subfigure[$\xi_{1}$]{\includegraphics[width = 0.5\textwidth]{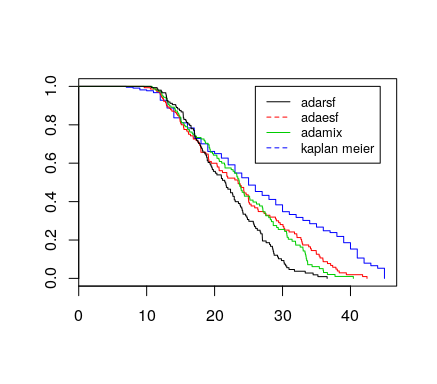}}  
 \subfigure[$\xi_{2}$]{\includegraphics[width = 0.5\textwidth]{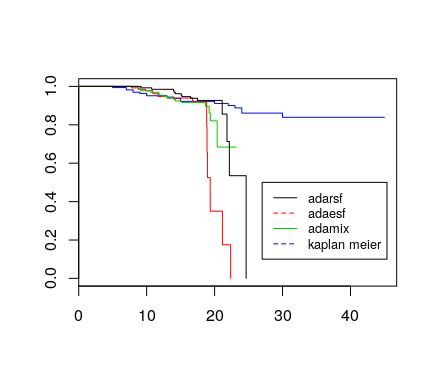}} \\
\caption{Mean of mode in PBC Data: Survival Function based on Kaplan Meier and different variations of Survival Trees \label{mean_mode_pbc}}
\end{center}
\end{figure}

\begin{figure}
\begin{center}
 \subfigure[$\xi_{1}$]{\includegraphics[width = 0.5\textwidth]{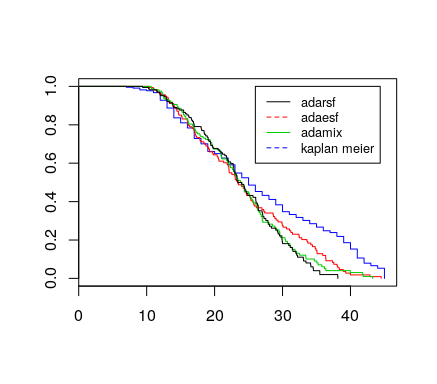}}  
 \subfigure[$\xi_{2}$]{\includegraphics[width = 0.5\textwidth]{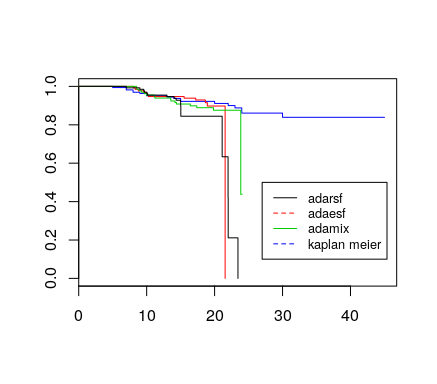}} \\
\caption{Mapped mean of mode for PBC Data : Survival Function based on Kaplan Meier and different variations of Survival Trees \label{map_mean_mode_pbc}}
\end{center}
\end{figure}

This dataset is available in R package survival.  Originally data is collected from the Mayo Clinic trial in primary biliary cirrhosis (PBC) of the liver, conducted between 1974 and 1984. A total of 424 PBC patients, referred to Mayo Clinic during that ten-year interval, met eligibility criteria for the randomized placebo controlled trial of the drug D-penicillamine.  The first 312 cases in the data set participated in the randomized trial and contain largely complete data. The additional 112 cases did not participate in the clinical trial, but consented to have basic measurements recorded and to be followed for survival. Six of those cases were lost to follow-up shortly after diagnosis, so the data here are on an additional 106 cases as well as the 312 randomized participants.  Here we plot two cause specific survival functions (transplant and death) based on two competing causes ignoring the missing observations.  Figure-\ref{mean_mode_pbc} and Figure-\ref{map_mean_mode_pbc} show the results obtained by mean of mode and mapped mean of mode respectively for both the causes. RMSE and system run time is calculated for each method for different competing risk.  The results for mean of mode method are available in Table-\ref{PBC-death-mean-mode}, Table-\ref{PBC-trans-mean-mode} whereas results for mapped mean of modes are available at Table-\ref{PBC-death-mapped-mean-modes} and Table-\ref{PBC-trans-mapped-mean-modes} respectively. 

\textbf{Important Observation :}  From both the figures we observe all the methods work quite well for PBC death case, whereas none of them works properly towards the tail for PBC transplant case.  One of the reason could be number of data set for transplant case is much lower as compared to death case.  Also PBC transplant may carry cure rate type of model and censored case with a very small data. Separate analysis is needed to address cure rate type of set up. In PBC death case, performance of ADA-ESF is better than other methods when Mean of mode is considered.  However ADA-MIX performs better both in RMSE as well as running time when Mapped mean of mode is considered. Overall, performance of ADA-ESF and ADA-MIX is closer. 

\begin{table}
        \centering
        \begin{tabular}{ |c|c|c|c| }
            \hline
            \multicolumn{4}{|c|}{RMSE}\\
            \hline
             & train & test & Running Time (sec)\\ 
            \hline
             ADA-ESF & 5.99 & 5.58 & 0.99\\ 
             ADA-MIX & 4.43 & 6.20 & 0.95\\ 
             ADA-RSF & 6.53 & 8.90 & 0.84\\ 
             \hline
        \end{tabular}
        \caption{PBC death case : Mean of modes - RMSE in 10 iterations each with 10 trees \label{PBC-death-mean-mode}}
\end{table}

\begin{table}
        \centering
        \begin{tabular}{ |c|c|c|c|}
            \hline
            \multicolumn{4}{|c|}{RMSE}\\
            \hline
             & train &test & Running Time (sec)\\ 
            \hline
             ADA-ESF & 10.88 & 11.28 & 0.47\\ 
             ADA-MIX & 9.62 & 11.35 & 0.44\\ 
             ADA-RSF & 9.60 & 9.68 & 0.47\\ 
             \hline
        \end{tabular}
        \caption{PBC transplant case : Mean of modes - RMSE in 10 iterations each with 10 trees \label{PBC-trans-mean-mode}}
\end{table}

\begin{table}
        \centering
        \begin{tabular}{ |c|c|c|c| }
            \hline
            \multicolumn{4}{|c|}{RMSE}\\
            \hline
             & train &test & Running Time (sec)\\ 
            \hline
             ADA-ESF & 4.32 & 6.36 & 1.07\\ 
             ADA-MIX & 4.58 & 5.64 & 0.95\\ 
             ADA-RSF & 5.48 & 7.07 & 4.71\\ 
             \hline
        \end{tabular}
        \caption{PBC death case : Mapped mean of modes - RMSE in 10 iterations each with 10 trees \label{PBC-death-mapped-mean-modes}}        
\end{table}

\begin{table}
        \centering
        \begin{tabular}{ |c|c|c|c| }
            \hline
            \multicolumn{4}{|c|}{RMSE}\\
            \hline
             & train &test & Running Time (sec)\\ 
            \hline
             ADA-ESF & 11.71 & 12.50 & 0.55\\ 
             ADA-MIX & 11.47 & 9.75 & 0.52\\ 
             ADA-RSF & 8.92 & 10.41 & 2.55\\ 
             \hline
        \end{tabular}
        \caption{PBC transplant case : Mapped mean of modes - RMSE in 10 iterations each with 10 trees \label{PBC-trans-mapped-mean-modes}}        
\end{table}

\section{Conclusion}  Predicting conditional survival function given the data set is a difficult problem.  In this paper we propose different variations of adaboost to predict the survival function.  We implement the same for three different data set. One of the data is very high dimensional in nature, integrates gene expression as covariates which shows that the algorithm is scalable.  The data set examples show that the proposed framework works for censoring and competing risk set up too.  In our algorithm we use log-rank statistic as split criteria.  However we can use other split criteria and compare the result.  From the above data analysis we see that it is safer to use ADA-ESF to obtain the best/near best RMSE and system run time.  Issues like variable importance, missing value, integration of pathway information etc are needed to be explored.  The work is on progress.

\nocite{IshwaranGerdsKogalurMooreGangeLau:2014} \nocite{VezhnevetsVezhnevets:2005} \nocite{KalbfleischPrentice:2011} \nocite{IshwaranKogalur:2007} \nocite{Segal:1988}  \nocite{KannaoGuha:2015}

\bibliographystyle{imsart-nameyear}
\bibliography{ada_cancer}

\end{document}